\documentclass[prb,twocolumn,floatfix]{revtex4}
\usepackage{graphicx}
\newcommand{\pdag}{{\phantom{\dagger}}}
\newcommand{\vc}[1]{{\bf #1}}
\def\k{{\bf{k}}}

\begin{document}
\title{From Slater to Mott-Heisenberg physics: The antiferromagnetic phase of the Hubbard model}
\author{Th.\ Pruschke}
\author{R.\ Zitzler}
\affiliation{Center for electronic correlations and magnetism,
Theoretical Physics III, Institute for Physics, University of Augsburg, 
86135 Augsburg, Germany}

\begin{abstract}
We study the optical conductivity of the one-band Hubbard model in the N\'eel state
at half filling at $T=0$ using the dynamical mean-field theory. For small values
of the Coulomb parameter clear signatures of a Slater insulator expected from a
weak-coupling theory are found, while the strongly correlated system can be well
described in terms of a Mott-Heisenberg picture. However, in contrast to the paramagnet,
we do not find any evidence for a transition between these two limiting cases but rather a smooth 
crossover as a function of the Coulomb interaction.
\end{abstract}
\pacs{}
\maketitle              

\section{Introduction}
The microscopic description of magnetism and metal-insulator transitions constitutes
one of the major research activities in modern solid state theory. For example, transition
metal compounds like V$_2$O$_3$, LaTiO$_3$, NiS$_{2-x}$Se$_{x}$ or the cuprates
show metal-insulator transitions and magnetic order depending on composition, pressure
or other control parameters.\cite{imada98} One interesting and controversial
question concerns the description of the optical properties of these materials,\cite{slater,mott,hubbard}
in particular whether the fundamental physics is governed by the broken translational symmetry
e.g.\ in the N\'eel state or rather by
correlations,\cite{thomas,loidl} i.e.\ the formation of
so-called Hubbard bands with an energy gap of the order of the relevant Coulomb repulsion.

\begin{figure}[htb]
\begin{center}
\includegraphics[width=0.45\textwidth,clip]{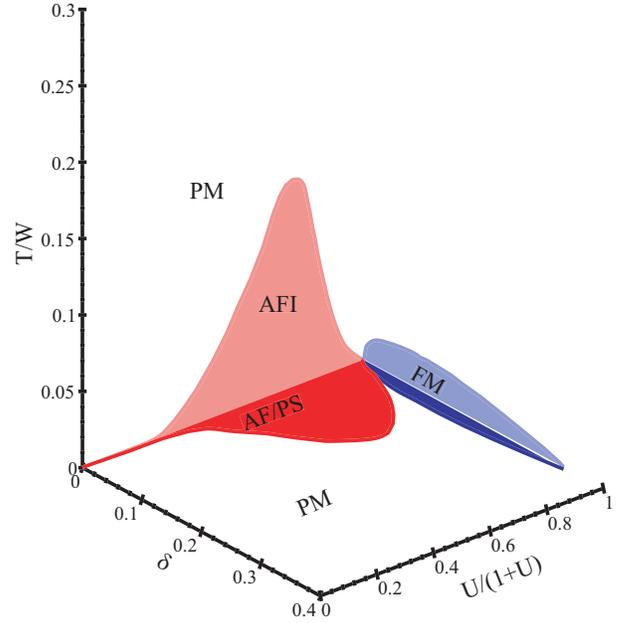}
\end{center}
\caption[]{Schematic DMFT phase diagram of the Hubbard model with
  nearest neighbor hopping on a simple cubic lattice. PM denotes the paramagnetic metal,
AFI the antiferromagnetic insulator, FM
the ferromagnetic metal and AF/PS phase separated antiferromagnetism.\label{fig:hmpd}}
\end{figure}

The simplest model showing both magnetism and a correlation induced metal-insulator
transition (MIT) is the one-band Hubbard model\cite{hubbard}
\begin{equation}
\label{equ:hubbard}
H=-\sum_{i,j,\sigma}t_{ij}
c^{\dagger}_{i\sigma}c^{\phantom{\dagger}}_{j\sigma}
+\frac{U}{2}\sum_{i\sigma}n_{i\sigma}n_{i\bar{\sigma}}\;\;,
\end{equation}
where $c^{(\dagger)}_{i\sigma}$ annihilates (creates) an electron at site $i$ with
spin $\sigma$,
$n_{i\sigma}=c^{\dagger}_{i\sigma}c^{\phantom{\dagger}}_{i\sigma}$,
$t_{ij}$ denotes the hopping amplitude between sites $i$ and $j$ and
$U$ is the
local Coulomb repulsion. Usually, one ignores longer range hopping processes
and concentrates on nearest neighbor hopping only.
Considerable progress in understanding the physics of this simple but nevertheless
non-trivial model has been achieved in the last decade through the development of
the dynamical mean-field theory (DMFT) \cite{mv,pradv,rmp}.
In particular, the phase diagram for
the Hubbard model on a simple cubic lattice with nearest-neighbor hopping
is very well understood \cite{pradv,rmp,jazph}. 
The major results are compiled in the schematic
phase diagram in Fig.~\ref{fig:hmpd}.
At half filling the physics is dominated by an antiferromagnetic insulating
phase (AFI) for all $U>0$. 
For finite doping, the antiferromagnetic phase persists up to a critical
doping $\delta_c$ \cite{jazph,zitz1} and in addition shows phase separation \cite{pvd,zitz1}.
For very large values of $U$ the antiferromagnetic phase is replaced
by a small region of Nagaoka type ferromagnetism.\cite{nagoka,oberm}

An appealing property of the DMFT is the possibility to calculate transport quantities in 
a very simple fashion. Due to the local nature of the theory, vertex corrections to the
leading particle-hole bubble of the current-current correlation function
vanish identically,\cite{khur90,pradv} i.e.\ one needs to calculate the
bare bubble only. This has been extensively used to study the optical
conductivity and various other
transport properties in the paramagnetic phase of the Hubbard model.\cite{pradv,rmp,raman}
On the other hand, up to now a comparable investigation of the optical properties of symmetry
broken phases, in particular the N\'eel state at half filling, has not been performed. However, such an
investigation is interesting for several reasons. First, the insulating phase in
real materials is in many cases accompanied by magnetic or orbital ordering,
typically of the N\'eel type. To what
extent the model (\ref{equ:hubbard}) can describe the optical properties of ordered insulating
phases has up to now not been studied in detail. Second, it is well-known that the restriction of
the Hubbard model to the paramagnetic state at half filling shows a metal-insulator transition\cite{rmp,jazph,bulprl,bulcosvol}
at a finite critical $U_c>0$ which is of first order.\cite{rmp,bulprl}
It might be argued
that for the N\'eel state a similar situation can occur. At small $U$ a weak
coupling theory is expected to give accurate results, leading to a band or Slater insulator\cite{slater}
due to the doubled unit cell in the N\'eel state.
At large $U$, on the other hand, the Hubbard model is known to reduce to an effective Heisenberg model\cite{fulde}
with localized moments from the onset. It is an open question whether these two limits are linked
continously or via a phase transition at some finite
value of the Coulomb interaction $U$.

The paper is organized as follows: In the next section the derivation of an expression for the optical
conductivity in the N\'eel state is presented. The results obtained for the optical conductivity
of the Hubbard model with nearest neighbor hopping on a simple cubic lattice at half filling are presented and discussed
in section III.  A conclusion and outlook finish the paper in section IV.

\section{Optical conductivity for the N\'eel state in DMFT}
\begin{figure}[htb]
\mbox{}
\begin{center}
\includegraphics[width=0.45\textwidth,clip]{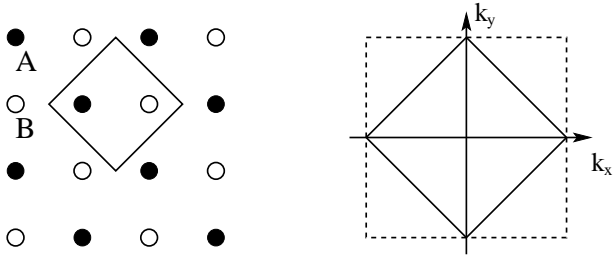}
\end{center}
\caption[]{Left: Schematic view of the $AB$ sublattice decomposition suitable
for the treatment of the N\'eel state. Right: Magnetic Brillouin zone (MBZ,
1.\ BZ of the N\'eel state).
\label{fig:AB}}
\end{figure}

In the N\'eel state, the DMFT equations have to be modified
to account for two inequivalent sublattices $A$ and $B$ (see Fig.~\ref{fig:AB},
left panel) with
self-energies $\Sigma^A \neq \Sigma^B$.\cite{rmp,brami}
To this end, we
introduce operators $a^{(\dagger)}_{i\sigma}$ and
$b^{(\dagger)}_{i\sigma}$ which act on sublattice $A$ and $B$
respectively. In the case of nearest-neighbor hopping only, the
kinetic part of the Hamiltonian (\ref{equ:hubbard}) can then be written as
\begin{displaymath}
 H_t=-t\sum_{\langle i,j \rangle}\sum_{\sigma} \left(
 a^{\dagger}_{i\sigma}b^{\phantom{\dagger}}_{j\sigma} +
 b^{\dagger}_{j\sigma}a^{\phantom{\dagger}}_{i\sigma} \right)\;\;.
\end{displaymath}
After Fourier transforming this expression we obtain
\begin{displaymath}
  H_t=\sum_{\sigma}{\sum_{\k}}^\prime
  \Psi^{\dagger}_{\k\sigma} \left( \begin{array}{cc} 0 & \epsilon_\k \\
  \epsilon_\k & 0 \end{array}
  \right) \Psi^{\phantom{\dagger}}_{\k\sigma}\;\;,
\end{displaymath}
where we introduced the spinors
\begin{displaymath}
  \Psi^{\dagger}_{\k\sigma} = \left( a^\dagger_{\k\sigma}\;,\;
  b^\dagger_{\k\sigma} \right) \;\;,\;\;
  \Psi^{\phantom{\dagger}}_{\k\sigma} = \left( \begin{array}{c}
  a^\pdag_{\k\sigma} \\ b^\pdag_{\k\sigma} \end{array} \right)
\end{displaymath}
and $\epsilon_\k$ is the dispersion on the
bipartite lattice. The prime on the sum indicates that the summation
is over all values of $\k$ in the {\em magnetic}\/ Brillouin zone (MBZ)
(see Fig.~\ref{fig:AB}, right panel).
Within this notation, the
Green function becomes a matrix in the two sublattices,
\begin{equation}
  \label{equ:gfmatrix}
  G_{\k\sigma}(z)=\left(\begin{array}{c@{\ \ }c}
  \zeta^{A}_\sigma & -\epsilon_\k\\
  -\epsilon_\k & \zeta^{B}_\sigma 
\end{array}\right)^{-1}\;\;,
\end{equation}
where $\zeta^{A/B}_\sigma=z+\mu-\Sigma^{A/B}_\sigma$. From now on we
employ the symmetry $\zeta^A_\sigma = \zeta^B_{\bar \sigma} \equiv
\zeta_\sigma$ of the N\'eel state and drop the indices $A/B$.
Using this formalism, the current operator is given by
\begin{displaymath}
  \vc j=e\sum_{\sigma}{\sum_\k}^\prime
  \Psi^{\dagger}_{\k\sigma} \left( \begin{array}{cc} 0 & \vc v_\k \\
  \vc v_\k & 0 \end{array}
  \right) \Psi^{\phantom{\dagger}}_{\k\sigma}
\end{displaymath}
with $\vc v_\k = \nabla_\k \epsilon_\k$ as usual.
If we consider a lattice for which the conductivity tensor is
diagonal,
the elements $\sigma_{ii} \equiv \sigma$ can be calculated from
($D$ is the spatial dimension of the lattice)
\begin{displaymath}
  D \cdot \sigma (\omega) = \Re e \frac{1}{i\omega} \sum_{l=1}^{D}
  \langle\!\langle j_l;j_l \rangle\!\rangle _{\omega+i\delta}
\end{displaymath}
with the current-current correlation function
\begin{eqnarray*}
  \lefteqn{\langle\!\langle j_l;j_l \rangle\!\rangle _{i\nu}= e^2
  \sum\limits_{\sigma\sigma^\prime} {\sum\limits_{\vc k \vc
  k^\prime}}^\prime v_\k^l v_{\k^\prime}^l} \\
  &\times& \langle\!\langle a_{\vc k\sigma}^\dagger b_{\vc
  k\sigma}^\pdag + b_{\vc k\sigma}^\dagger a_{\vc
  k\sigma}^\pdag;a_{\vc k^\prime\sigma^\prime}^\dagger b_{\vc
  k^\prime\sigma^\prime}^\pdag+b_{\vc k^\prime\sigma^\prime}^\dagger
  a_{\vc k^\prime\sigma^\prime}^\pdag \rangle\!\rangle _{i\nu}\;\;.
\end{eqnarray*}
Again, due to the symmetry of the lattice, the index $l$ can be
dropped. The most important simplification arises from the locality of
two-particle self-energies within the DMFT.\cite{khur90,brami,prub} Note that
in the present formulation the proper locality of the two-particle self-energies
is still ensured, because in the DMFT as defined by equ.~(\ref{equ:gfmatrix})
no dynamical correlations between the $A$ and $B$ sublattices are introduced.
In analogy to
the paramagnetic case this allows us to carry out the $\k$ sums in
diagrams containing two-particle self-energy insertions independently at
each vertex. Since the single particle propagators only depend on $\k$
through the even function $\epsilon_\k$ and the $v_\k$ are
of odd parity, the sum over their product vanishes. As a result, we
obtain the exact expression for the current-current correlation function
in the DMFT,
\begin{eqnarray*}
  \lefteqn{\langle\!\langle j;j \rangle\!\rangle
  _{i\nu}=-\frac{e^2}{\beta}\sum\limits_{\omega_n} {\sum\limits_\sigma
  \sum\limits_{\vc k}}^\prime v_{\vc k}^2} \\
  &\times& \left[ \langle\!\langle a_{\vc k\sigma}^\pdag;a_{\vc
  k\sigma}^\dagger \rangle\!\rangle _{i\omega_n+i\nu} \,
  \langle\!\langle b_{\vc k\sigma}^\pdag;b_{\vc k\sigma}^\dagger
  \rangle\!\rangle _{i\omega_n} \right. \\
  &+& \left. \langle\!\langle b_{\vc k\sigma}^\pdag;b_{\vc k\sigma}^\dagger
  \rangle\!\rangle _{i\omega_n+i\nu} \, \langle\!\langle a_{\vc
  k\sigma}^\pdag ;a_{\vc k\sigma}^\dagger \rangle\!\rangle
  _{i\omega_n} \right. \\
  &+& \left. \langle\!\langle b_{\vc k\sigma}^\pdag ;a_{\vc k\sigma}^\dagger
  \rangle\!\rangle _{i\omega_n+i\nu} \, \langle\!\langle b_{\vc
  k\sigma}^\pdag ;a_{\vc k\sigma}^\dagger \rangle\!\rangle
  _{i\omega_n} \right. \\
  &+& \left. \langle\!\langle a_{\vc k\sigma}^\pdag ;b_{\vc k\sigma}^\dagger
  \rangle\!\rangle _{i\omega_n+i\nu} \, \langle\!\langle a_{\vc
  k\sigma}^\pdag ;b_{\vc k\sigma}^\dagger \rangle\!\rangle
  _{i\omega_n} \right]\;\;.
\end{eqnarray*}
In terms of the Green function matrix elements in (\ref{equ:gfmatrix})
we can rewrite this as
\begin{eqnarray*}
  \lefteqn{\langle\!\langle j;j \rangle\!\rangle
  _{i\nu}=-\frac{e^2}{\beta}\sum\limits_{\omega_n} {\sum\limits_\sigma
  \sum\limits_{\vc k}}^\prime v_{\vc k}^2} \\
  &\times& \left[ G_{\k\sigma}^{AA}(i\omega_n+i\nu)\,
  G_{\k\sigma}^{BB}(i\omega_n) \right. \\
  &+& \left. G_{\k\sigma}^{BB}(i\omega_n+i\nu)\,
  G_{\k\sigma}^{AA}(i\omega_n) \right. \\
  &+& \left. G_{\k\sigma}^{BA}(i\omega_n+i\nu)\,
  G_{\k\sigma}^{BA}(i\omega_n) \right. \\
  &+& \left. G_{\k\sigma}^{AB}(i\omega_n+i\nu)\,
  G_{\k\sigma}^{AB}(i\omega_n) \right]
\end{eqnarray*}
where
\begin{displaymath}
  G_{\k\sigma}^{AA}(z)=\frac{\zeta_{\bar\sigma}}{\zeta_\sigma
  \zeta_{\bar\sigma} - \epsilon_\k^2}\;\;,\;\;
  G_{\k\sigma}^{BB}(z)=\frac{\zeta_{\sigma}}{\zeta_\sigma
  \zeta_{\bar\sigma} - \epsilon_\k^2}
\end{displaymath}
and
\begin{displaymath}
  G_{\k\sigma}^{BA}(z)=G_{\k\sigma}^{AB}(z) =
  \frac{\epsilon_\k}{\zeta_\sigma \zeta_{\bar\sigma} - \epsilon_\k^2}\;\;.
\end{displaymath}
Next, we convert the $\k$ sum into an energy integral by introducing
the average squared velocity, 
\begin{equation}\label{equ:vsqrav}
\langle v^2\rangle_\epsilon:=\frac{1}{D\cdot N}{\sum_{\k}}'v_{\k}^2\delta(\epsilon
-\epsilon_{\k})\;\;.
\end{equation}
Making furthermore use of the
spectral representation of the Green functions, the frequency sum can
be evaluated in a straightforward way and finally we obtain for the
conductivity
\begin{displaymath}
  \sigma(\omega) = c \sum\limits_\sigma
  \int\limits_{-\infty}^0 \! d\epsilon\,
  \langle v^2\rangle_\epsilon \int\limits_{-\infty}^\infty \! d\omega^\prime\,
  \frac{f(\omega^\prime)-f(\omega^\prime \! + \omega)}{\omega}
\end{displaymath}
\begin{equation}
  \label{equ:optcond}
  \times \left[ A_\sigma(\epsilon,\omega^\prime)
  A_{\bar{\sigma}}(\epsilon,\omega^\prime \! + \omega)
  + B_\sigma(\epsilon,\omega^\prime)
  B_\sigma(\epsilon,\omega^\prime \! + \omega) \right]
\end{equation}
with
\begin{displaymath}
  A_\sigma(\epsilon,\omega)= -\frac{1}{\pi} \Im m
  \,G^{AA}_\sigma(\epsilon,\omega+i\delta)
\end{displaymath}
and
\begin{displaymath}
  B_\sigma(\epsilon,\omega)= -\frac{1}{\pi} \Im m
  \,G^{AB}_\sigma(\epsilon,\omega+i\delta) \;\;.
\end{displaymath}
Here $f(\omega)$ is the Fermi function and $c$ collects various
constants. Note that the form (\ref{equ:optcond}) is reminiscent of the
results found in the case of superconductivity, which is discussed at
length e.g.\ in the book by Mahan.\cite{mahan} Consequently, one can expect
to obtain similar features from the evaluation of
(\ref{equ:optcond}).

In order to proceed with the calculation,
it is necessary to specify the actual lattice structure and the corresponding
non-interacting dispersion in equation
(\ref{equ:vsqrav}). For the hypercubic lattice,\cite{prub}
$\langle v^2\rangle_\epsilon\propto\rho^{(0)}(\epsilon)$ is a simple Gaussian,
and the integration over $\epsilon$ can be performed
analytically. For the details of this calculation see Appendix A.

\section{Results}
In the following we present results for the optical properties of the Hubbard model
on a simple hypercubic lattice with nearest neighbor hopping at half filling and $T=0$ in the DMFT. 
The hopping matrix element is chosen as $t=t^\ast/\sqrt{4D}$, which ensures the correct scaling of
the kinetic energy in the limit $D\to\infty$.\cite{mv}
As energy unit it is convenient to use the bandwidth $W$ of the system at $U=0$.
Since the Gaussian density of states (DOS) of the simple hypercubic lattice in the limit of infinite spatial dimensions
has no real band edges, we take $W=4t^\ast$ as a reasonable value. Note that for this choice of
$W$, (i) the spectral weight of the Gaussian is exhausted by $99$\% between
$\omega=\pm W/2$ and (ii) the paramagnetic metal-insulator transition
will occur at $U_c\approx 4t^\ast=W$.\cite{jazph}
The effective quantum impurity model of the DMFT is solved using Wilson's Numerical Renormalization Group
method (NRG),\cite{nrg} suitably extended for dynamical quantities and spin-polarization.\cite{bulprhew,hof01} The
calculations were performed with a discretization parameter $\Lambda=2$, keeping $800$ states. Dynamical
quantities were calculated with a Gaussian logarithmic broadening of $0.6$. Occasional checks with $1600$
states or smaller $\Lambda$ showed sufficient robustness of the results.

\subsection{Single-particle properties}
\begin{figure}[htb]
\mbox{}
\begin{center}
\includegraphics[width=0.45\textwidth,clip]{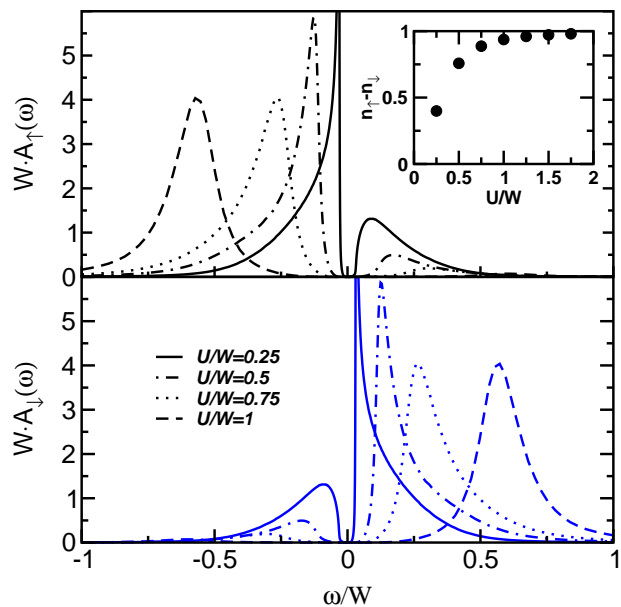}
\end{center}
\caption[]{Spin-resolved DOS for the antiferromagnetically ordered phase of the half-filled Hubbard model
for different values of $U/W$. The inset shows the magnetization as function of $U/W$. 
\label{fig:DOSofU}}
\end{figure}
Before discussing the optical conductivity calculated from (\ref{equ:optcond}), it is helpful to review the 
single particle properties. The spin resolved one-particle DOS calculated at $T=0$ for different values of $U$
shows the expected insulating behavior with a clear gap at the Fermi energy for all $U>0$. In particular for
small values of $U\ll W$, the DOS shows nicely developed nesting
singularities at the gap edges, which qualitatively
follow the predictions of a weak coupling theory.\cite{zitz1} With increasing $U$ these features get
more and more smeared out, and for $U\agt W$ the spectra resemble those of the Mott insulator.\cite{pradv}
Note that the appearance of a gap in the DOS is of course accompanied
by a vanishing
imaginary part of the one-particle self-energy in this region.
Neither the development of the DOS nor the magnetization as a function of $U$ shown in the inset to
Fig.~\ref{fig:DOSofU} provide any evidence as to whether the limits $U\ll W$ and $U\agt W$ will be linked
smoothly or by some kind of transition.

\subsection{Spin dynamics}
Another interesting quantity is the dynamical magnetic susceptibility, whose
low-energy behavior gives further insight into possible differences in spin
and charge dynamics. In principle, it is also possible to calculate this
quantity as a function of wave vector $\vc q$ within the DMFT.\cite{rmp,jazph} However,
this requires the calculation of the local irreducible particle-hole self-energy,\cite{brami} which is
\begin{figure}[htb]
\mbox{}
\begin{center}
\includegraphics[width=0.45\textwidth,clip]{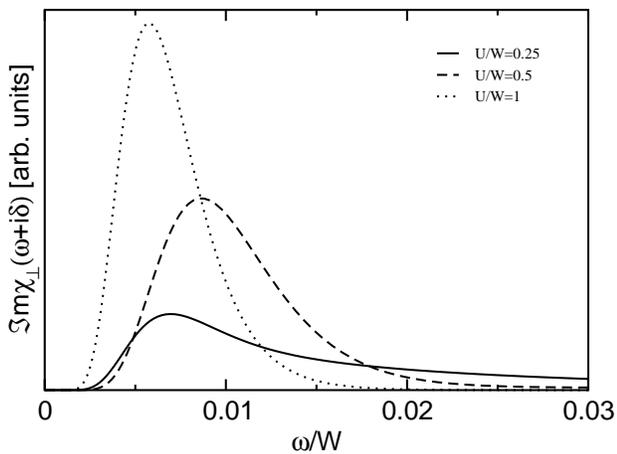}
\end{center}
\caption[]{Imaginary part of the local transverse magnetic
  susceptibility as a function of $\omega/W$ 
for $U/W=0.25$, $0.5$ and $1$, which shows a well-defined gap $\Delta_s$ as $\omega\to0$. Only the part for
$\omega>0$ is shown. Note that $\Delta_s$ first increases with increasing $U$, but eventually
decreases again.\label{fig:spingap}}
\end{figure}
presently not possible within the NRG. Nevertheless, for the current
investigation, a reasonable approximation 
can be obtained from
the local magnetic susceptibility, 
$$
\chi_\bot(z)=\frac{1}{N}\sum\limits_{\vc q} \chi_\bot(\vc q,z)\;\;.
$$ 
Since the ground state
of our system is the N\'eel state, spin excitations require a minimum
excitation energy,  the spin gap $\Delta_s$, which conventionally is read off
$\Im m\chi_\bot(\vc Q,\omega+i\delta)$ evaluated at the antiferromagnetic
wave vector $\vc Q=(\pi,\pi,\ldots)$.
However, the gaps at other $\vc q$ vectors will be equal to or larger than $\Delta_s$. Thus, even after summing over all
$\vc q$-values, the size of the gap in $\Im m\chi_\bot(\omega+i\delta)$ will
be determined by $\Delta_s$.
The quantity $\Im m\chi_\bot(\omega+i\delta)$ on the other hand can
easily be calculated from the NRG once the DMFT has converged. The results for three
typical values of $U/W$ are shown in Fig.~\ref{fig:spingap}, displaying a nice spin gap $\Delta_s$ as $\omega\to0$.
Evidently, the value of $\Delta_s$ first increases with increasing $U$ but then decreases again, as is to be
expected from the mapping of the Hubbard model to an antiferromagnetic Heisenberg model with $J\propto 1/U$ at
large $U$. 
From the calculated $\Im m\chi_\bot(\omega+i\delta)$ one can directly extract
the values for $\Delta_s(U)$.
The results will be discussed below together
with the charge gap obtained from the optical conductivity (see Fig.~\ref{fig:spin_charge_gap}).

\subsection{Optical conductivity and optical gap}
\begin{figure}[htb]
\mbox{}
\begin{center}
\includegraphics[width=0.45\textwidth,clip]{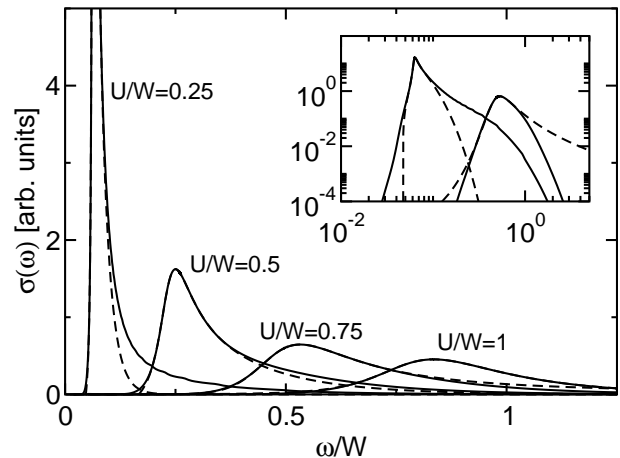}
\end{center}
\caption[]{Optical conductivity of the half-filled Hubbard model in the N\'eel
  phase for $T=0$ as a function of $U$. The full lines
represent the calculated data, the dashed lines a fit with the function $\omega\cdot\sigma(\omega)=
\Im
m\left\{e^{i\phi}\left(\omega-\omega_0+i\gamma\right)^{-\alpha}\right\}$
(see text). The inset shows the curves for $U/W=0.25$ and $U/W=0.75$
using a logarithmic scaling.
\label{fig:optcond}}
\end{figure}

The optical conductivity resulting from the spectra in Fig.~\ref{fig:DOSofU} is shown in Fig.~\ref{fig:optcond}. 
Apparently, the overall behavior seen in the DOS has its counter part in $\sigma(\omega)$. For small values of $U$,
one finds a threshold behavior with a singularity, whereas for large $U$ the optical conductivity closely resembles the
one found in the paramagnetic insulator.\cite{jazph} Obviously, there are at least two interesting features in
$\sigma(\omega)$. First, the behavior of $\sigma(\omega)$ in the vicinity of the maximum and second the actual
value of the optical gap, i.e.\ the energy at which $\sigma(\omega)=0$.

In order to address the first point we adopt the following line of reasoning. In the Hartree limit,
\begin{figure}[htb]
\mbox{}
\begin{center}
\includegraphics[width=0.45\textwidth,clip]{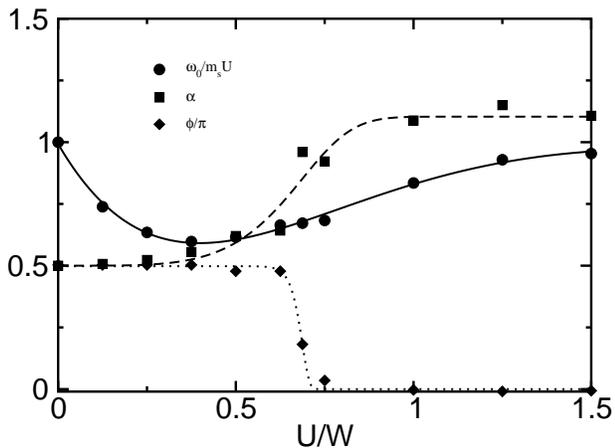}
\end{center}
\caption[]{Dependence of the fit parameters $\omega_0$, $\alpha$ and $\phi$ in (\ref{equ:fitfunction}) on $U$. The
lines are meant as guides to the eye. Note the rather well defined change in $(\alpha,\phi)$ from
$(\alpha,\phi)=(1/2,\pi/2)$ to $(\alpha,\phi)=(1,0)$ around $U/W=0.75$.
\label{fig:fitparms}}
\end{figure}
i.e.\ without an imaginary part of the self-energy, an approximate evaluation of (\ref{equ:optcond}) yields
$$
\omega\cdot\sigma(\omega)\propto \frac{\Theta(\omega-2\Delta_0)}{\sqrt{\omega-2\Delta_0}}
$$
with $\Delta_0=Um_s/2$ and $m_s=\langle n_\uparrow-n_\downarrow\rangle$. Since this behavior is goverened by square root
singularities in the integrand in (\ref{equ:optcond}) (see e.g.\ the explicit
formula derived in the appendix),
it is reasonable to assume that for a finite imaginary part of the self-energy the above singularity will become an algebraic function
\begin{equation}\label{equ:fitfunction}
\omega\cdot\sigma(\omega)\propto \Im m\left\{\frac{e^{i\phi}}{\left(\omega-\omega_0+i\gamma\right)^\alpha}\right\}
\end{equation}
with a general exponent $\alpha$. The quantity $\gamma$ approximately cares for the finite imaginary part introduced by the
one-particle self-energy and $\phi$ allows for a more complex mixing of real and imaginary parts in the integral
(\ref{equ:optcond}). The function (\ref{equ:fitfunction}) describes the behavior of $\sigma(\omega)$ in the
vicinity of the maximum very nicely for all values of $U$ (see
dashed lines in Fig.~\ref{fig:optcond}); note that from the inset to Fig.~\ref{fig:optcond}
it is evident that for small $U$ this algebraic form has the tendency to overestimate the optical gap, while
at large $U$ it is clearly underestimated.

Let us now turn to the behavior of the parameters $\omega_0$,
$\alpha$ and $\phi$ shown in Fig.~\ref{fig:fitparms}. As $U\to0$, we expect that $\omega_0=2\Delta_0=Um_s$, $\alpha=1/2$
and $\phi=\pi/2$, i.e.\ $\omega\cdot\sigma(\omega)\propto \Re e\left(\omega-\omega_0+i\delta\right)^{-1/2}=
\Theta(\omega-\omega_0)/\sqrt{\omega-\omega_0}$. 
We indeed find the anticipated square-root singularity; however, even for
small $U/W$, the value of $\omega_0$
significantly deviates from the Hartree value, being systematically
smaller but obviously approaching it as $U\to0$. 

For values $U>W$, the behavior of $\omega\cdot\sigma(\omega)$ is best described by a Lorentian, which becomes apparent
from the values of $\alpha$ and $\phi$ obtained in this region, viz $\alpha\approx1$ and $\phi=0$, meaning 
$\omega\cdot\sigma(\omega)\propto \Im m\left(\omega-\omega_0+i\gamma\right)^{-1}\propto 1/\left((\omega-\omega_0)^2+\gamma^2\right)$. 
In addition, the results for $\omega_0$ together with $m_s\approx1$ indicate that $\omega_0\approx U$, in agreement
with the predictions of the Mott-Hubbard picture.\cite{rmp}

The behavior of the optical gap $\Delta_c$,
\begin{figure}[htb]
\mbox{}
\begin{center}
\includegraphics[width=0.45\textwidth,clip]{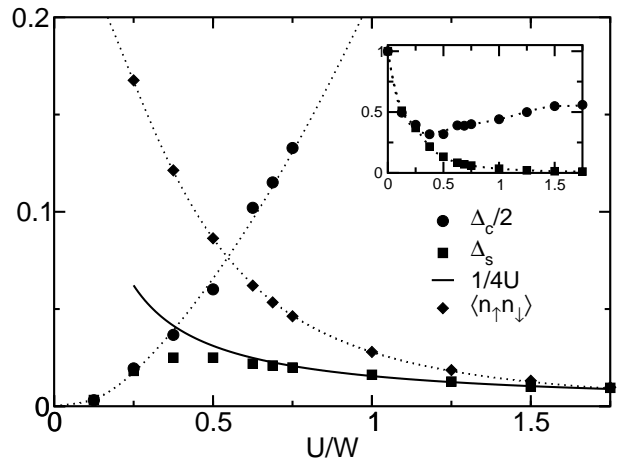}
\end{center}
\caption[]{Optical gap $\Delta_c/2$ (circles), spin gap $\Delta_s$ (squares) and double occupancy
(diamonds) as function of $U$. The inset shows the gaps scaled with $Um_s/2$.
Dotted lines are meant as guides to the eye.  
For small $U$ both charge and spin gap are
identical, while for large $U$ we find $\Delta_c\propto U$ (see inset) and $\Delta_s\propto 1/U$
(full line in main panel).\label{fig:spin_charge_gap}}
\end{figure}
together with the spin gap $\Delta_s$ obtained from $\Im m\chi_\bot(\omega+i\delta)$ and the double
occupancy $\langle n_\uparrow n_\downarrow\rangle$, is displayed in
Fig.~\ref{fig:spin_charge_gap}.
The details of the method used to obtain $\Delta_c$ are discussed in appendix 
B.

For small $U/W$, the optical gap is exactly twice as large as the spin gap and,
as becomes apparent from the inset, approaches the Hartree value $m_sU$ as
$U\to0$. 
Again {\em both} quantities deviate systematically and by the same amount from the
expected Hartree values even for the smallest $U$. Thus, even for $U/W<<1$
correlation effects are important and significantly modify the predictions from Hartree theory.\cite{pvd,mouk}
For large $U$, on the other hand, we find $\Delta_c\propto U-W$, consistent with Mott-Hubbard localized states;
furthermore, $\Delta_s\propto1/U$ as expected from the mean-field theory of the
Heisenberg model with a $J\propto 1/U$. 

We find, however, no evidence that the Slater limit at $U/W\to0$ and
the Mott-Heisenberg limit at $U/W\to\infty$  are separated by some kind of
phase transition. All results, including the variation of
$\langle n_\uparrow n_\downarrow\rangle$ seen in
Fig.~\ref{fig:spin_charge_gap}, rather indicate that a smooth crossover takes
place for $U/W \approx 3/4$.

\section{Conclusion and outlook}
While in the paramagnetic phase of the Hubbard model at half filling, when artificially extended to $T=0$,
a true phase transition from a correlated metal to a Mott-Hubbard insulator at $U_c\approx W$ has been
established, the situation in the physically more relevant N\'eel state has not been investigated
in similar detail up to now. As a first step into this direction the properties in the ground state
of the Hubbard model at half filling with particle-hole symmetry have been discussed. We did confirm that the
physical properties at small and large values of the Coulomb parameter $U$ can be well described within a
Slater and Mott-Heisenberg picture, respectively. In contrast to the paramagnetic Mott-Hubbard MIT we
could not find any solid evidence for a similar transition in the N\'eel state; our data rather suggest a
smooth crossover, which occurs at a value $U\alt W$. Even the double
occupancy, which in the case of the paramagnetic MIT is
an indicator of a phase transition, does not show any sign of an anomaly here.

A novel quantity we discussed
was the spin gap, which we extracted from the local transverse spin susceptibility. The general
behavior and size agree very well with exactly known limits. This shows that at least in cases where a well defined
spin gap exists that becomes minimal at special points in the Brillouin zone, even the inspection of purely local
dynamical susceptibilities can be sufficient.

There are, however, still several unanswered questions. First, the analytic form of the optical conductivity
close to the optical gap and the precise value of this gap could not be obtained at present due to numerical
problems when evaluating the integral (\ref{equ:optcond}). Especially for a more quantitative comparison with
experiment this has to be improved in future work. Second, comparison with the data for V$_2$O$_3$ 
from ref.~\onlinecite{thomas}
\begin{figure}[htb]
\mbox{}
\begin{center}
\includegraphics[width=0.45\textwidth,clip]{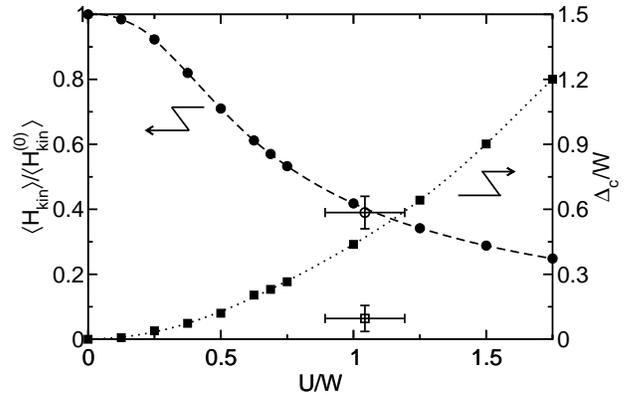}
\end{center}
\caption[]{Kinetic energy scaled to its value at $U=0$ (circles and left scale) and full optical gap $\Delta_c/W$ 
(squares and right scale) vs. $U/W$. The open circle and square with
errorbars represent values for the
scaled kinetic energy and $\Delta_c/W$ respectively, extracted from ref.~\onlinecite{thomas} for the sample of V$_2$O$_3$ with
$T_N\approx50$K. Note that in ref.~\onlinecite{thomas} $D=W/2$ and $\Delta=\Delta_c/2$.\label{fig:H_kin}}
\end{figure}
show a nice agreement for the kinetic energy (obtained using
the optical sum rule), but fail completely concerning the size of the optical
gap. This is shown
in Fig.~\ref{fig:H_kin}, where we plot the kinetic energy scaled to its value for $U=0$ (circles and left scale) and
$\Delta_c/W$ (squares and right scale) vs.\ $U/W$. The open circle and square represent data for the kinetic energy
and charge gap, respectively, extracted from ref.~\onlinecite{thomas} for a 
V$_2$O$_3$ sample with $T_N\approx50$K. Obviously,
with the present model, one largely overestimates the optical gap at intermediate values of $U$.

Of course the present investigation did concentrate on the simplest situation, viz a system with perfect particle-hole
symmetry. In reality, electron hopping beyond nearest neighbors will destroy antiferromagnetism at small values of $U$
and consequently lead to different gaps at intermediate values of $U$. On the other hand, the gaps at large $U$ are
controlled by Mott-Hubbard physics and will most likely change only little. A similar line of argument has in
fact been invoked in ref.~\onlinecite{thomas}, too. A recent study of the
properties of the magnetically frustrated Hubbard model indicates that
this scenario is indeed very likely.\cite{zitz03} Evidently, a further detailed
investigation of the optical properties in
the N\'eel state, in particular with more realistic band structures, is
necessary and surely highly interesting. Work along this line is in progress.

\begin{acknowledgments}
We acknowledge useful conversations with
D.\ Vollhardt,
T.P.\ Deveraux,
G.\ Uhrig,
M.\ Potthoff,
S.\ Kehrein
and
R.\ Bulla.
This work was supported by the DFG through the collaborative research center SFB 484, the
Leibniz Computer center and the Computer
center of the Max-Planck-Gesellschaft in Garching.
\end{acknowledgments}

\appendix

\section{Further evaluation of (\ref{equ:optcond})}
In this appendix we present details of the further evaluation of the energy
integrals in equation (\ref{equ:optcond}) for the hypercubic lattice in the limit
$D\to\infty$. In that case the density of states becomes a Gaussian, i.e.\ we need to
calculate the two integrals
\begin{displaymath}
  \int_{-\infty}^{\infty} d\epsilon \,e^{-\epsilon^2} A_\sigma(\epsilon,\omega^\prime)
  A_{\bar{\sigma}}(\epsilon,\omega^\prime \! + \omega)
\end{displaymath}
and
\begin{displaymath}
  \int_{-\infty}^{\infty} d\epsilon \,e^{-\epsilon^2} B_\sigma(\epsilon,\omega^\prime)
  B_\sigma(\epsilon,\omega^\prime \! + \omega) \;\;.
\end{displaymath}
It is now convenient to split up the spectral functions into two parts, i.e.\
\begin{displaymath}
A_\sigma(\epsilon, \omega) = A_\sigma^-(\epsilon, \omega) + A_\sigma^+(\epsilon, \omega)
\end{displaymath}
with
\begin{displaymath}
A_\sigma^\pm(\epsilon, \omega) = -\frac{1}{2\pi} \Im m  \frac{\zeta_{\bar\sigma}}{\sqrt{\zeta_\sigma \zeta_{\bar\sigma}}} \left( \frac{1}{
\sqrt{\zeta_\sigma \zeta_{\bar\sigma}}
    \pm \epsilon} \right) 
\end{displaymath}
and $\zeta_\sigma$ as defined after equ.\ (\ref{equ:gfmatrix}).
In the same way we write
\begin{displaymath}
B_\sigma(\epsilon, \omega) = B_\sigma^-(\epsilon, \omega) - B_\sigma^+(\epsilon, \omega)
\end{displaymath}
where now
\begin{displaymath}
B_\sigma^\pm(\epsilon, \omega) = -\frac{1}{2\pi} \Im m  \frac{1}{
\sqrt{\zeta_\sigma \zeta_{\bar\sigma}}
    \pm \epsilon} \;\;.
\end{displaymath}
Using this notation and  collecting equivalent terms, it can easily be verified that the following four
integrals need to be evaluated:
\begin{displaymath}
  I_1=2\int_{-\infty}^{\infty} d\epsilon \,e^{-\epsilon^2} B_\sigma^-(\epsilon,\omega^\prime)
  B_\sigma^-(\epsilon,\omega^\prime \! + \omega)  \;\;,
\end{displaymath}
\begin{displaymath}
  I_2=-2\int_{-\infty}^{\infty} d\epsilon \,e^{-\epsilon^2} B_\sigma^-(\epsilon,\omega^\prime)
  B_\sigma^+(\epsilon,\omega^\prime \! + \omega) \;\;,
\end{displaymath}
\begin{displaymath}
  I_3=2\int_{-\infty}^{\infty} d\epsilon \,e^{-\epsilon^2} A_\sigma^-(\epsilon,\omega^\prime)
  A_{\bar{\sigma}}^-(\epsilon,\omega^\prime \! + \omega)
\end{displaymath}
and
\begin{displaymath}
  I_4=2\int_{-\infty}^{\infty} d\epsilon \,e^{-\epsilon^2} A_\sigma^-(\epsilon,\omega^\prime)
  A_{\bar{\sigma}}^+(\epsilon,\omega^\prime \! + \omega)\;\;.
\end{displaymath}
The further evaluation will be demonstrated for the first
term. Using the notation
\begin{displaymath}
  \alpha = \left. \sqrt{\zeta_\sigma \zeta_{\bar\sigma}}
  \,\right|_{\omega^\prime +
  i\delta} \;\;{\rm and} \;\; \beta = \left. \sqrt{\zeta_\sigma
  \zeta_{\bar\sigma}}
  \,\right|_{\omega^\prime \! + \omega + i\delta}
\end{displaymath}
we may write
\begin{eqnarray*}
  \lefteqn{B_\sigma^-(\epsilon,\omega^\prime)
  B_\sigma^-(\epsilon,\omega^\prime \! + \omega)=} \\
  &&-\frac{1}{4\pi^2} \left[ \left( \frac{1}{\overline\alpha - \epsilon} -
  \frac{1}{\alpha - \epsilon} \right) \left( \frac{1}{\overline\beta - \epsilon} -
  \frac{1}{\beta - \epsilon} \right) \right] \;\;,
\end{eqnarray*}
where the bar above a term denotes complex conjugation.
The terms inside the brackets can be expanded further to yield
\begin{eqnarray*}
  \lefteqn{\left[ -\frac{1}{\overline\alpha-\overline\beta}
  \left(\frac{1}{\overline\alpha -\epsilon}-\frac{1}{\overline\beta-\epsilon}\right)
  +\frac{1}{\overline\alpha-\beta}
  \left(\frac{1}{\overline\alpha -\epsilon}-\frac{1}{\beta-\epsilon}\right)
  \right.} \\
  &&\left. \!\!\!\!\!+\frac{1}{\alpha-\overline\beta}
  \left(\frac{1}{\alpha -\epsilon}-\frac{1}{\overline\beta-\epsilon}\right)
  -\frac{1}{\alpha-\beta}
  \left(\frac{1}{\alpha -\epsilon}-\frac{1}{\beta-\epsilon}\right)
  \right]\,.
\end{eqnarray*}
If we introduce the Faddeeva function
\begin{displaymath}
  w(z)=\frac{i}{\pi} \int_{-\infty}^{\infty} dt\,
  \frac{e^{-t^2}}{z-t}=e^{-z^2} {\rm erfc}\,(-iz)
\end{displaymath}
for complex arguments $z$ with $\Im m z > 0$, we obtain
\begin{eqnarray*}
  I_1&=&\frac{1}{2\pi i} \left[ \frac{w(\alpha)-w(\beta)}{\alpha-\beta}
  -\frac{\overline{w(\alpha)}-\overline{w(\beta)}}{\overline\alpha-\overline\beta}
  \right. \\
  &&\quad\quad\left. -\frac{w(\alpha)+\overline{w(\beta)}}{\alpha-\overline\beta}
  +\frac{w(\alpha)+\overline{w(\beta)}}{\overline\alpha-\beta}
  \right] \;\;,
\end{eqnarray*}
where we have made use of the relation $w(-\overline z)=\overline{w(z)}$. 
The remaining three contributions can be obtained in a similar fashion. Finally, 
combining complex conjugate expressions, we arrive
at the following result for the four integrals:
\begin{equation}
  \label{eq:i3}
  I_1=\frac{1}{\pi} \, \Im m \left(\frac{w(\alpha)-w(\beta)}{\alpha
  -\beta} - \frac{w(\alpha)+\overline{w(\beta)}}{\alpha-\overline\beta} \right)
\end{equation}
\begin{equation}
  \label{eq:i4}
  I_2=-\frac{1}{\pi} \, \Im m \left(\frac{w(\alpha)-\overline{w(\beta)}}{\alpha
  +\overline\beta} - \frac{w(\alpha)+w(\beta)}{\alpha+\beta} \right)
\end{equation}
\begin{equation}
  \label{eq:i1}
  I_3=\frac{1}{\pi} \, \Im m \left(\gamma\delta
  \frac{w(\alpha)-w(\beta)}{\alpha -\beta} - \gamma\overline\delta
  \frac{w(\alpha)+\overline{w(\beta)}}{\alpha-\overline\beta} \right)
\end{equation}
\begin{equation}
  \label{eq:i2}
  I_4=\frac{1}{\pi} \, \Im m \left(\gamma\overline\delta
  \frac{w(\alpha)-\overline{w(\beta)}}{\alpha+\overline\beta} - \gamma\delta
  \frac{w(\alpha)+w(\beta)}{\alpha+\beta} \right)
\end{equation}
where we have introduced
\begin{displaymath}
  \gamma = \left. \frac{\zeta_{\bar\sigma}}{\sqrt{\zeta_\sigma
  \zeta_{\bar\sigma}}} \,\right|_{\omega^\prime +
  i\delta} \;\;{\rm and}\;\; \delta =
  \left. \frac{\zeta_{\sigma}}{\sqrt{\zeta_\sigma
  \zeta_{\bar\sigma}}} \,\right|_{\omega^\prime \! + \omega + i\delta} \;\;.
\end{displaymath}
A further analytical evaluation of the remaining integration over $\omega^\prime$ in
equation (\ref{equ:optcond}) using eqs.\ (\ref{eq:i3}) -- (\ref{eq:i2}) is possible
only for $\zeta_\sigma\to\omega-\sigma\Delta_0+i\delta$. In this case,
the square-roots appearing in the functions $\alpha$ and $\beta$ lead to a
typical threshold behavior of the form\cite{mahan}
$$
\omega\cdot\sigma(\omega)\propto\frac{\Theta(\omega-2\Delta_0)}{\sqrt{\omega-2\Delta_0}}\;\;.
$$

The appearence of this threshold singularity also shows that a further numerical evaluation
of the remaining integral over $\omega^\prime$ in equation (\ref{equ:optcond}) will
become problematic in regions where the imaginary part of the one-particle self-energy
becomes small, because the integrand will develop a strongly singular behavior.
In particular, this makes a precise numerical evaluation of the
optical conductivity near the threshold impossible.

\section{The optical gap}
While the definition of the optical gap is straight forward, the extraction
of numbers from the numerical data appears to be rather problematic for
two simple reasons. First, the spectra calculated with NRG have an unavoidable
intrinsic broadening, which becomes especially severe for the Hubbard bands at
larger values of $U/W$. Second, as $\omega\ll\omega_0$, the imaginary part of
the one-particle self-energy becomes negligible, and the singular structure of
the integrand (\ref{eq:i3}) -- (\ref{eq:i2}) entering
into (\ref{equ:optcond}) makes additional broadening necessary to
allow for a stable numerical integration. Together both effects very efficiently
mask the true $\omega$-dependence close to the optical gap, in particular
for larger $U/W$.

\begin{figure}[htb]
\mbox{}
\begin{center}
\includegraphics[width=0.45\textwidth,clip]{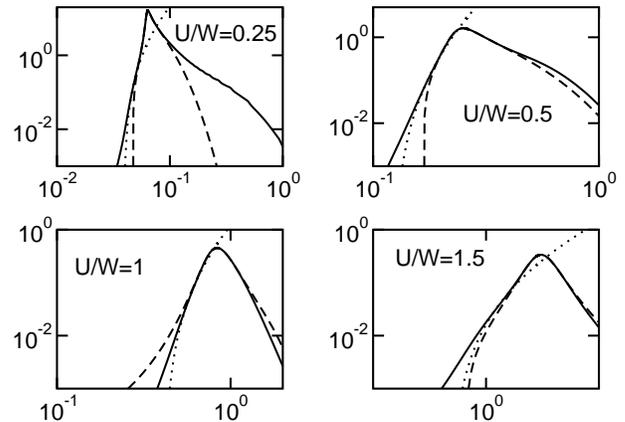}
\end{center}
\caption[]{Results for the fits to $\sigma(\omega)$ (full lines) with function (\ref{equ:fitfunction}) 
for $\omega\approx\omega_0$ (dashed lines) and 
$\Theta(\omega-2\Delta_0)\cdot\left(\omega-2\Delta_0\right)^{5/2}$ in the low-$\omega$ region (dotted lines)
for several values of $U/W$.\label{fig:lowwfit}}
\end{figure}

In order to nevertheless have an unambiguous working procedure that allows
to extract a reasonable approximation to the true optical gap from our numerical data, we {\em postulate} that
$\omega\cdot\sigma(\omega)\propto \Theta(\omega-2\Delta_0)\cdot\left(\omega-2\Delta_0\right)^\alpha$
for $\omega$ in the region where (\ref{equ:fitfunction}) starts to deviate substantially from the data
and {\em choose} a minimal $\alpha$ such that it produces a reasonable fit {\em for all values of $U$}
(see Fig.~\ref{fig:lowwfit} for selected results). We find $\alpha=5/2$\cite{comment} and an optical gap $\Delta_c$ 
which is consistent with the spin gap $\Delta_s$ as $U\to0$. The
good agreement of these two differently calculated quantities (see inset
to Fig.~\ref{fig:spin_charge_gap})
also serves as an a posteriori check for the procedure used to determine $\Delta_c$.
In view of a possible comparison to
experimental results\cite{thomas} this situation is, of course, not satisfying. For this purpose a more thorough
and possibly analytical evaluation of $\sigma(\omega)$ close to $\Delta_c$ would be desirable. 
Unfortunately, the complicated form of the integrals
in (\ref{equ:optcond}) so far have allowed for an analytical
evaluation only in the Hartree limit.


\begin{thebibliography}{99}
\bibitem{imada98} M.~Imada, A.~Fujimori, and Y.~Tokura, Rev.\ Mod.\ Phys.\ {\bf 70},
1039 (1998).
\bibitem{slater} J.C.~Slater, Phys.\ Rev.\ {\bf 82}, 538 (1951).
\bibitem{mott} N.F.~Mott, Philos.\ Mag.\ {\bf 6}, 287 (1961).
\bibitem{hubbard} J.~Hubbard, Proc.\ R.\ Soc.\ London A{\bf 276}, 238 (1963);
M.C.~Gutzwiller, Phys.\ Rev.\ Lett. {\bf 10}, 59 (1963);
J.~Kanamori, Prog.\ Theor.\ Phys. {\bf 30}, 275 (1963).
\bibitem{thomas} G.A.~Thomas, D.H.~Rapkine, S.A.~Carter, A.J.~Millis, T.F.~Rosenbaum, P.~Metcalf and
J.M.~Honig, Phys.\ Rev.\ Lett.\ {\bf 73}, 1529 (1994).
\bibitem{loidl}, A.~Loidl, private communication.
\bibitem{mv} W.\ Metzner und D.\ Vollhardt, Phys.\ Rev.\ Lett. {\bf 62},
324 (1989).
\bibitem{pradv} T.~Pruschke, M.~Jarrell and J.K.~Freericks, Adv. 
Phys.~{\bf 42}, 187 (1995);
\bibitem{rmp} A.~Georges, G. Kotliar, W. Krauth and M.J. Rozenberg, 
Rev.~Mod.~Phys.~{\bf 68}, 13 (1996).
\bibitem{jazph} M.~Jarrell and Th.~Pruschke, Z.\ Phys.\ B{\bf 90},
  187 (1993).
\bibitem{zitz1} R.~Zitzler, Th.~Pruschke, R.~Bulla, Eur.\ Phys.\ J.\ B
  {\bf 27}, 473 (2002).
\bibitem{pvd} P.G.J.~van~Dongen, Phys.\ Rev.\ Lett. {\bf 67}, 757 (1991);
Phys.\ Rev.\ B{\bf 50}, 14016 (1994).
\bibitem{nagoka} Y.~Nagaoka, Phys.\ Rev.\ {\bf 147}, 392 (1966).
\bibitem{oberm} Th.~Obermeier, Th.~Pruschke and J.~Keller, Phys.\ Rev.\ 
B{\bf 56},
R8479 (1997).
\bibitem{khur90} A.~Khurana, Phys.\ Rev.\ Lett.\ {\bf 64}, 1990 (1990).
\bibitem{raman} J.K.~Freericks, T.P.~Deveraux, and R.~Bulla, Phys.\ Rev.\ B{\bf 64}, 233114 (2001);
J.K.~Freericks, T.P.~Deveraux, R.~Bulla and Th.~Pruschke, Phys.\ Rev.\ B {\bf 67}, 155102 (2003).
\bibitem{bulprl} R.~Bulla, Phys.\ Rev.\ Lett. {\bf 83}, 136 (1999). 
\bibitem{bulcosvol} R.~Bulla, T.A.~Costi, D.~Vollhardt Phys.\ Rev.\
  B{\bf 64}, 045103 (2001).
\bibitem{fulde} P.~Fulde, {\it Electron correlations in molecules and solids},
           Springer, Berlin (1995).
\bibitem{brami} U.~Brandt und C.~Mielsch, Z.\ Phys.\ B{\bf 82}, 37 (1991).
\bibitem{prub} Th.~Pruschke, D.L.~Cox, M.~Jarrell, Phys.\ Rev.\ B{47},
  3553 (1993).
\bibitem{mahan} G.~Mahan, {\it Many-Particle Physics}, Plenum Press,
  New York (1990).
\bibitem{nrg} K.G. Wilson, Rev. Mod. Phys. {\bf 47}, 773 (1975);
 H.R. Krishna-murthy, J.W. Wilkins, and K.G. Wilson,
  Phys. Rev. B {\bf 21}, 1003 (1980); {\it ibid.} {\bf 21}, 1044 (1980).
\bibitem{bulprhew} R.~Bulla,  A.C.~Hewson and Th.~Pruschke,
J.\ Phys.\ -- Condens.\ Matter {\bf 10}, 8365 (1998).
\bibitem{hof01} T.A.~Costi, Phys.\ Rev.\ Lett. {\bf 85},
  1504(2000); W.~Hofstetter, Phys.\ Rev.\ Lett. {\bf 85}, 1508 (2000).
\bibitem{mouk} S.~Moukouri and M.~Jarrell, Phys.\ Rev.\ Lett.\ {\bf 87},
167010 (2001).
\bibitem{zitz03} R.~Zitzler, N.~Tong, Th.~Pruschke, and R.~Bulla,
cond-mat/0308202.
\bibitem{comment} The value $\alpha=3/2$ used in
  ref.~\onlinecite{thomas} does not lead to a satisfying description.
\end{thebibliography}
\end{document}